\documentclass[aps,prb,twocolumn,superscriptaddress,showpacs]{revtex4}
\usepackage{graphicx}
\begin{document}


\title{Unconventional spin fluctuations in the hexagonal antiferromagnet YMnO$_3$}


\author{T. J. Sato}
    \email[]{tjsato@nist.gov}
    \altaffiliation{On leave from National Institute for Materials Science, Tsukuba 305-0047, Japan.}
\author{S. -H. Lee}
\affiliation{NIST Center for Neutron Research, National Institute of Standards and Technology, Gaithersburg, MD 20899}
\author{T. Katsufuji}
\affiliation{Department of Physics, Waseda University, Shinjuku-ku, Tokyo 169-8555, Japan}
\author{M. Masaki}
\affiliation{Department of Advanced Materials Science and Department of Applied Chemistry, University of Tokyo, Tokyo 113-8656, Japan}
\author{S. Park}
\affiliation{NIST Center for Neutron Research, National Institute of Standards and Technology, Gaithersburg, MD 20899}
\affiliation{\mbox{Department of Materials and Nuclear Engineering, University of Maryland, College Park, MD 20742}}
\author{J. R. D. Copley}
\affiliation{NIST Center for Neutron Research, National Institute of Standards and Technology, Gaithersburg, MD 20899}
\author{H. Takagi}
\affiliation{Department of Advanced Materials Science and Department of Applied Chemistry, University of Tokyo, Tokyo 113-8656, Japan}



\date{\today}

\begin{abstract}
We used inelastic neutron scattering to show that well below its
N\'{e}el temperature, $T_{\rm N}$, the two-dimensional (2D) XY nearly-triangular
antiferromagnet YMnO$_{3}$ has a prominent {\it central peak} associated with 2D antiferromagnetic fluctuations with a characteristic life time of 0.55(5) ps, coexisting with the conventional long-lived spin-waves. 
Existence of the two time scales suggests competition between the
N\'{e}el phase favored by weak interplane interactions, and the
Kosterlitz-Thouless phase intrinsic to the 2D $XY$ spin
system.
\end{abstract}


\maketitle


Geometrical frustration and low dimensionality are the two key concepts in the statistical physics that provide
unusual spin dynamics as well as phase transitions~\cite{ram01,lee02,jon87}.
The simplest realization of the two concepts is two-dimensional triangular lattice antiferromagnets (2DTLAFM).
A particular interest is placed on the $XY$ spin system (2D$XY$TLAFM), where its ground-state manifold has the continuous degeneracy associated with $U(1)$ global spin rotations, as well as the discrete Ising-like degeneracy due to $Z_2$ chirality configurations~\cite{kaw98}.
Because of the $U(1)$ symmetry, the well-known Kosterlitz-Thouless (KT) phase involving vortex binding~\cite{kos73} is expected at low temperatures.
Experimentally, little evidence can be found in the literature due to the lack of good model systems~\cite{col97}.

Hexagonal rare-earth manganites $R$MnO$_3$ ($R = $Y, Lu, and Sc;
space group $P6_3cm$) can be good candidates for the 2DTLAFMs. The
Mn$^{3+}$ ($S = 2$) ions form nearly triangular networks in $z =
0$ and $1/2$ layers, stacking with the ABAB sequence (see
Fig.~1)~\cite{mun00,kat01,kat02}. The layers are well separated by
$R$ and O ions, suggesting good 2D character in the $ab$ plane.
Their bulk susceptibility data show that despite strong
antiferromagnetic interactions the magnetic ions order long-range 
at much lower temperatures $T_{\rm N}$ than the magnetic energy scale inferred by the Curie-Weiss
temperatures $\Theta_{\rm CW}$ (for instance, $\Theta_{\rm CW} =
-705$~K and $T_{\rm N} = 70$~K for YMnO$_3$)~\cite{mun00,kat01,kat02}.
Previous powder
neutron diffraction studies showed that the spins at the lowest temperature formed the so-called 120$^{\circ}$ structure in the $ab$-plane coinciding with the
ground state for 2DTLAFMs, and the frozen moments, $\langle M \rangle$,
were reduced from the expected value for the fully polarized Mn$^{3+}$ ({\it e.g.}, $\langle M \rangle =
2.90(2)\mu_{\rm B} < gS\mu_{\rm B}$ for YMnO$_3$ ($S = 2$))~\cite{mun00,kat02,bie99}.
The reduction in $T_N$ and $\langle M \rangle$ is a signature of strong spin fluctuations
due to geometrical frustration and/or low dimensionality in the
systems. A broad peak was additionally observed at finite $Q$
around $T_{\rm N}$, indicative of strong short-range spin
correlations~\cite{bie99,kat02}. However, due to intrinsic
limitations of the powder diffraction technique, further
experimental studies, especially inelastic single crystal neutron
scattering measurements are necessary to understand the nature of
the spin excitations.

In this paper, we report for the first time inelastic neutron
scattering measurements on powder and single-crystal samples of the
hexagonal rare-earth manganite YMnO$_3$. We have found that
YMnO$_3$ is a good model system for the 2D$XY$TLAFM with weak
trimerization. Our most important finding is that in the N\'{e}el
phase there are fast 2D spin fluctuations with a characteristic
time scale of 0.55(5)~ps in addition to the conventional long-lived
spin wave excitations. The inverse of dynamic correlation length
associated with the fast 2D spin fluctuations has similar
$T$-dependence as that expected for the KT phase in a 2D $XY$ spin system, 
suggesting that the spin fluctuations are reminiscence of the KT phase.
The coexistence of the long-lived magnons and the fast 2D spin fluctuations also
suggests competition between
the N\'{e}el phase and the KT phase in this quasi-2D $XY$ spin
system.

\begin{figure}
 \includegraphics[scale=0.25, angle=-90]{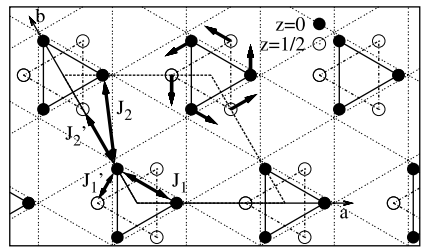}%
 \caption{\label{fig1} Schematic drawing of Mn positions and spin ordering in YMnO$_3$.
    Filled (open) circles represent Mn positions in the $z=0$ ($z = 1/2$) plane,
    whereas dotted parallelogram shows the magnetic unit cell that is identical to the chemical unit cell.
    Lattice constants are $a = 6.140$~\AA\ and $c = 11.393$~\AA.
    The lattice is weakly trimerized with the intra-trimer and inter-trimer Mn-Mn distances of 3.42~\AA\ and 3.62~\AA, respectively~\cite{kat01,kat02}, which is exaggerated in drawing.}
\end{figure}

A 50~g powder sample and a 2~g ($\phi 5$~mm $\times$ 22~mm) single crystal of YMnO$_3$ were
used in our neutron scattering measurements.
Methods of sample preparation were reported elsewhere~\cite{kat01}. Neutron scattering measurements were performed at the NIST Center for Neutron Research.
Powder experiments were performed at the Disk Chopper time-of-flight Spectrometer (DCS) using an incident energy of $E_{\rm i} = 15.46$~meV and single crystal experiments at the cold neutron triple-axis spectrometer SPINS and the thermal neutron triple-axis spectrometer BT9.
At SPINS, pyrolytic graphite (PG) 002 reflections were used for monochromator and analyzer, and a cooled Be filter was placed after the sample to eliminate higher-order contamination.
We used horizontal collimations of 80$'$-80$'$ and a final energy $E_{\rm f} = 5$~meV for most scans, while $E_{\rm f} = 2.6$~meV and 80$'$-40$'$ were used when better energy resolution was needed.
At BT9, the PG monochromator and analyzer were used with $E_{\rm f} = 14.7$~meV,
and a PG filter was used to get rid of higher order contamination.

\begin{figure}
 \includegraphics[scale=0.40, angle=0]{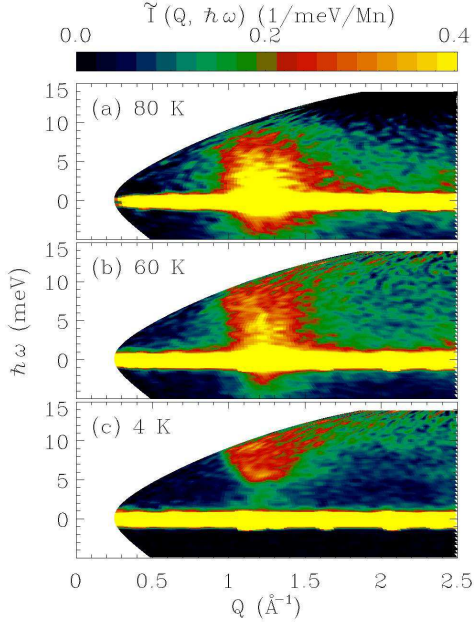}%
 \caption{\label{fig2} Color contour maps of the powder-averaged neutron scattering intensity versus magnitude of wave vector transfer $Q$ and energy transfer $\hbar\omega$ at three different temperatures spanning the phase transition at $T_{\rm N} = 70$ K.
}
 \end{figure}

Fig.~2 provides an overview of the inelastic neutron scattering intensity $\tilde{I}(Q,\omega)$ for the powder sample at three temperatures.
The powder averaged scattering intensity is related to the dynamic spin correlation function ${\cal S}^{\alpha \beta}(\vec{Q}, \omega)$ as~\cite{lov84},
\begin{equation}
\tilde{I}(Q,\omega ) = \int \frac{{\rm d}\Omega_{\hat{Q}}}{4\pi}
    |\frac{g}{2}F(Q)|^2 \sum_{\alpha\beta}(\delta_{\alpha\beta}-\hat{Q}_\alpha \hat{Q}_\beta )
    {\cal S}^{\alpha\beta}(\vec{Q},\omega ),
\end{equation}
where $F(Q)$ is the magnetic form factor for Mn$^{3+}$.
For $T > T_{\rm N}$, there is a cooperative paramagnetic continuum centered at $Q = 1.2$ \AA$^{-1}$ due to fluctuations of small AFM clusters, as is commonly found in geometrically frustrated AFMs~\cite{lee00}.
By integrating $\tilde{I}(Q,\omega )$ over $\hbar\omega$ and $Q$, we obtained the sum rule of $S(S+1) = 5.2(5)$ at 80~K, which is close to the expected value for dynamic Mn$^{3+} (S=2)$ ions.
This and the $Q$ dependence~\cite{phonon} tell us that the scattering is magnetic.
For $T < T_{\rm N}$, as the magnetic long range order develops, spectral weight at low energies gradually shifts to higher energies.
At $T = 4$~K there is strong scattering above $\hbar\omega \sim 5$~meV and weak scattering below.

\begin{figure}
    \includegraphics[scale=0.4, angle=-90]{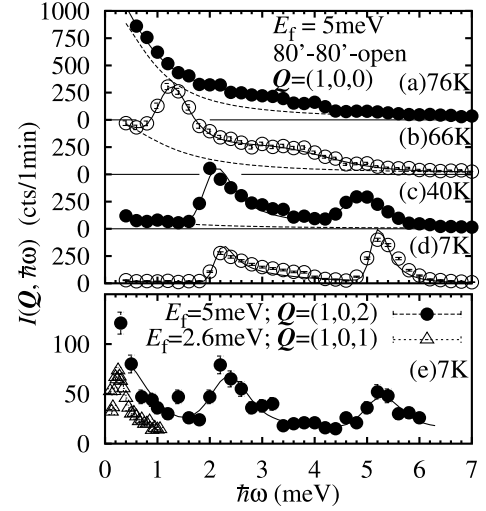}
 \caption{ (a-d) Constant $\vec{Q} =(1,0,0)$ scans at four different temperatures.
    (e) Constant-$\vec{Q}$ scans at $\vec{Q} = (1,0,2)$ and $\vec{Q} = (1,0,1)$ at $T = 7$~K.
    The scan at $\vec{Q} = (1, 0, 1)$ was taken under the higher-energy-resolution configuration with $E_{\rm f} = 2.6$~meV, and its unit is arbitrary.
    Solid lines are fits to Eq.~(\ref{fittingfunction}), whereas dashed lines represent the quasielastic part (see the text).
}
\end{figure}

Next, to obtain $\vec{Q}$-directional dependence of the magnetic excitations, we have performed single crystal inelastic scattering experiments.
Fig.~3 shows the representative constant-$\vec{Q}$ scans at the antiferromagnetic zone center $\Gamma$, namely, $\vec{Q} = (1,0,0)$ and equivalent positions.
For $T > T_{\rm N}$, the cooperative paramagnetic continuum appears as a quasielastic peak centered at $\hbar\omega = 0$~meV.
For $T < T_{\rm N}$, the quasielastic peak intensity decreases and two prominent magnon peaks develop at nonzero energies.
The energy values of the magnon peaks increase as $T$ decreases, becoming $\hbar\omega = 2.3$ and 5.3~meV at 7~K.
A constant $\vec{Q} = (1,0,1)$ scan with a better energy resolution revealed an additional mode at $\hbar\omega = 0.22$~meV (Fig.~3(e)).

We analyzed the observed spectra using the following scattering function with Lorentzians for the quasielastic and magnon peaks:
\begin{eqnarray}\label{fittingfunction}
\tilde{I}(\vec{Q}, \hbar\omega) &\propto& \hbar\omega[1 + n(\hbar\omega)]\left[ I_{\rm qel}\frac{\Gamma_{\rm qel}}{\Gamma_{\rm qel}^2 + \hbar\omega^2} \right. \nonumber \\
     &+& \left. \sum_{k} I_{\rm SW}^{k}\frac{\Gamma_{\rm SW}}{\Gamma_{\rm SW}^2 + (\hbar\omega - \hbar\omega_{k})^2}\right],
\end{eqnarray}
where $[1 + n(\hbar\omega)] = [1 -\exp(-\hbar\omega/k_{\rm B}T)]^{-1}$.
This function was convoluted with the instrumental resolution function to fit the observed spectra.

\begin{figure}
     \includegraphics[scale=0.3, angle=-90]{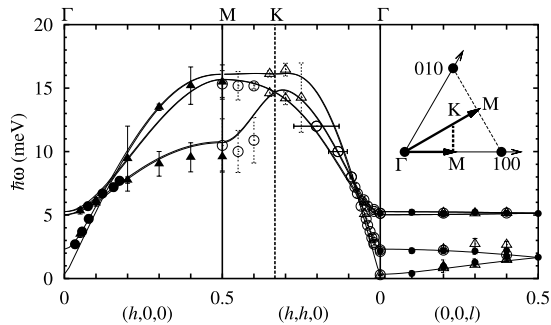}%
\caption{
 Spin wave dispersion relations along $(h,0,0), (h,h,0)$ and $(0,0,l)$.
    Circles and triangles represent the experimentally determined peak positions.
    (Different marks distinguish branches appearing from different magnetic Bragg positions.)
    Solid lines are the model dispersion relations explained in the text.    
    Inset: schematic drawing of scan directions in the 2D plane.    }
\end{figure}

Let us first discuss the magnon dispersion relations at $T = 7$~K $<< T_{\rm N}$. Fig.~4 shows the dispersion relations along a few high symmetry directions, obtained from several constant-$\vec{Q}$ and constant-$\hbar\omega$ scans.
To explain the observed dispersion relations, we introduce the following model spin Hamiltonian:
\begin{equation}
{\cal H} = -\sum_{<ij>} J_{ij} \vec{S}_i \cdot \vec{S}_j
    - D_1 \sum_{i} (S_i^z)^2
    - D_2 \sum_{i} (\vec{S}_i \cdot \vec{n}_i)^2,
\end{equation}
which consists of two inplane ($J_1$ and $J_2$) and two interplane ($J_1'$ and $J_2'$) interactions, and the easy-plane ($D_1$) and inplane easy-axis ($D_2$) anisotropies (see Fig.~1 for the definition of the interactions).
The anisotropy $D_2$, parallel to the spin directions ($\vec{n}_i = \langle \vec{S}_i \rangle/|\langle \vec{S}_i \rangle |$), is necessary to reproduce the small (0.22~meV) gap at the antiferromagnetic zone center, and is presumably due to the local structural distortion around Mn$^{3+}$.
The conditions $J_1' > 0$ and $J_1' > J_2'$ are necessary for the particular interplane stacking in YMnO$_3$ to be the ground state.

The model Hamiltonian is linearized using the Holstein-Primakoff approximation, and numerically diagonalized to obtain one-magnon dispersion relations using the standard equation-of-motion technique~\cite{whi65}.
Analytic expressions for the gap energies at the $\Gamma$ point were also derived assuming sufficiently small $D_2$, $J_1'$ and $J_2'$:
$\hbar \omega_1 \simeq 2 S \sqrt{- D_2 \lambda_1}, \hbar \omega_2 \simeq 2 S \sqrt{- D_2 \lambda_1 -2(J_1'-J_2')\lambda_1}, \hbar \omega_3 \simeq S \sqrt{2(D_1 \lambda_2 - D_2 \lambda_3 - 2D_1J_1')}$ and $\hbar \omega_4 \simeq S \sqrt{2(D_1 \lambda_2 - D_2 \lambda_3 - D_1(J_1' - 4J_2') - 2(J_1' - J_2')\lambda_2)}$ (from low to high energies), where $\lambda_1 = D_1 + (3/2)J_1 + 3J_2, \lambda_2 = (3/2)J_1 + 3J_2$ and $\lambda_3 = 2D_1 + (3/2)J_1 + 3J_2$.
Fitting the calculations to the data, we obtained $J_1 = -3.4(2)$~meV, $J_2 = -2.02(7)$~meV, $J_1' - J_2' = 0.014(2)$~meV, $D_1 = -0.28(1)$~meV and $D_2 = 0.0007(6)$~meV.
Solid lines in Fig.~4 represent the calculated dispersion relations with $J_2^{'}=0$.
The good agreement confirms the validity of the model Hamiltonian.
In the above, we could only determine the difference $J_1' - J_2'$ for interplane interactions.
From the analytic expressions we see that $\hbar\omega_3$ and $\hbar\omega_4$ must be accurately determined in order to obtain $J_1'$ and $J_2'$ separately.
However, this was impossible since they appear as one peak at $\hbar\omega = 5.3$~meV in Fig.~3(d) or 3(e) due to the insufficient energy resolution at high energies.
Since the splitting between $\hbar\omega_3$ and $\hbar\omega_4$ becomes sensitive to $J_1'$ (or $J_2'$) at $\vec{Q} = (1.05,0,0)$, we performed a constant-$\vec{Q}$ scan at this $\vec{Q}$ and found an almost resolution-limited peak at $\hbar\omega = 5.4$~meV.
This requires the splitting to be less than the energy resolution $\Delta E = 0.5$~meV, and consequently an upper limit of 0.08~meV is obtained for $J_1'$ and $J_2'$.
Hence, the interplane interactions are at most 2.4~\% of the inplane interaction $J_1$, evidencing the good two-dimensionality. One may note that $J_1 \sim J_2$, which makes YMnO$_3$ rather closer to the ideal TLAFM than a system of weakly coupled trimers.
Our $J_1 \simeq -3.4$~meV is one order of magnitude smaller than $J$ deduced in a recent Raman scattering study~\cite{tak02}.
They obtained $J \sim -140$~cm$^{-1}$ ($ \sim -17$~meV) by assigning a broad peak appearing at 1800~cm$^{-1}$ ($ \sim 220$~meV) to two-magnon scattering.
However, our results clearly show that the peak cannot be due to the two-magnon process because the band width of the one-magnon branch is only about 16~meV. Their broad peak at 220 meV must be vibrational or electronic in origin rather than magnetic.

\begin{figure}
    \includegraphics[scale=0.3, angle=-90]{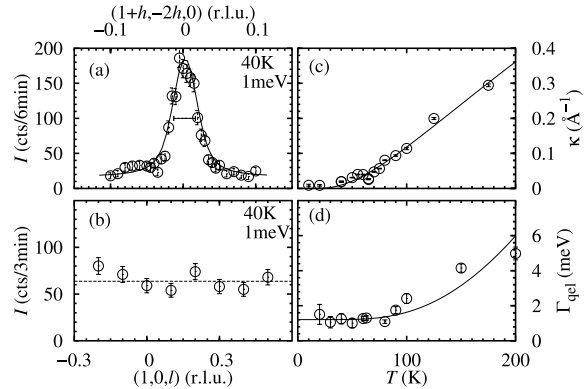}

 \caption{
    Constant-$\hbar\omega = 1$~meV scan (a) along the $(1+h, -2h, 0)$ direction and (b) along the $(1, 0, l)$ direction at $T = 40$~K.
    The horizontal bar in (a) represents the instrumental resolution.
    $T$-dependence of (c) the intrinsic peak width, $\kappa$, obtained from the constant-$\hbar\omega = 1$~meV scans the along $(1+h, -2h, 0)$ direction, and
    (d) the relaxation rate $\Gamma_{\rm qel}$ obtained from the constant-$\vec{Q} = (1,0,0)$ scans shown in Fig.~3. Lines are explained in the text.}
\end{figure}

Now let us turn to the low energy quasielastic continuum observed
below $T_{\rm N}$, clearly seen in Fig.~3(b) and 3(c). For $T = 7$
K $<< T_N$, Fig.~3(e) shows a $\hbar\omega=0.55$ meV mode at
$\vec{Q}=(1,0,l)$ with $l\neq 0$ which is due to inplane transverse
spin fluctuations. The inplane transverse fluctuations cannot,
however, appear at $\vec{Q}=(1,0,0)$ because the polarization factor
in Eq.~(1) vanishes for the ordered spin structure in YMnO$_3$. 
Note that such a mode does not show in Fig.~3(d).
Therefore we rule out the inplane transverse spin fluctuations as
the origin of the quasielastic continuum existing in the N\'{e}el
phase. In order to understand the continuum, we performed constant
$\hbar\omega=1$ meV around $(1,0,0)$ at several temperatures and
along different $\vec{Q}$-directions. Shown in Fig.~5(a) and 5(b)
are representative scans at $T=40$~K. A nearly
resolution-limited peak is seen along the inplane $(1+h, -2h, 0)$ direction
whereas the intensity is independent of $l$ perpendicular to the plane. These
indicate that the quasielastic component is purely 2D in nature, well
localized at the 2D antiferromagnetic zone center. Fig.~5(c) shows
the temperature dependence of the intrinsic peak width along the
$(1+h, -2h, 0)$ direction. For $T > T_{\rm N}$ the width decreases
almost linearly, whereas it becomes nearly resolution-limited
below $T_{\rm N}$, indicating a large inplane correlation length
at low temperatures. The energy width $\Gamma_{\rm qel}$ of the
quasielastic peak is also shown in Fig.~5(d). $\Gamma_{\rm qel}$
decreases as $T$ decreases down to $T_N$ and saturates to a value
of $\Gamma_{\rm qel} \simeq 1.2(1)$~meV below $T_N$. It is
surprising that the fast spin fluctuations with the characteristic
time scale of $\tau_{\rm qel} = \hbar/\Gamma_{\rm qel} \sim
0.55(5)$~ps coexist with the long-lived spin-waves in the N\'{e}el
phase.

What is the origin of the fast 2D fluctuations in the N\'{e}el phase? 
Recently, a similar quasielastic peak, called {\it central peak}, has been
found in numerical simulation studies on 2DXYTLAFMs~\cite{nho02}.
Theoretically, such a central peak has been commonly seen in 2D $XY$ spin systems, triangular or non-triangular, and is related to the vortex dynamics intrinsic to the KT phase~\cite{cos96}.
We, thus, fitted our $\kappa$ and $\Gamma_{\rm qel}$ to the phenomenological functions:
$\kappa = \kappa_0 \exp(-b/\sqrt{\tau})$ and 
$\Gamma = \Gamma_0 + A \pi (\Lambda/\hbar) {\rm e}^{-2b/\sqrt{\tau}}[(\sqrt{2} - 1)(\ln(k_{\rm B}T_{\rm KT}/\Lambda)/2 + b/\sqrt{\tau})]^{1/2}$, where $\tau = (T-T_{\rm KT})/T_{\rm KT}$ and $\Lambda = JS^2a^2\kappa_0^2/4$.
Here if $A = 1$ and $\Gamma_0 = 0$, $\kappa$ and $\Gamma$ reduce to the analytical expressions for the 2D $XY$ square lattice system~\cite{mer89}. 
The best fit (solid lines in Fig.~5(c) and 5(d)) was obtained with $T_{\rm KT} = 11(10)$~K, $b = 10(4)$, $\kappa_0 = 4(1)$~\AA$^{-1}$, $A = 0.07(1)$ and $\Gamma_0 = 1.2(1)$~meV. 
The fit reproduces $\kappa$ and $\Gamma$ well for the entire temperature range, suggesting that the quasielastic peak is reminiscence of the vortex dynamics.
Coexistence of the magnons and quasielastic peak suggests competition between the N\'{e}el phase favored by the weak interplane interactions, and the KT phase intrinsic to the 2D $XY$ spin system at low temperatures.
It remains to be seen whether or not the prefactor $A$ being smaller than 1 and the nonzero $\Gamma_0$ for the relaxation rate are intrinsic to the 2DXYLTAFM or are due to the competition between the two phases~\cite{sac00}. 

In summary, using inelastic neutron scattering measurements on a
single crystal of the hexagonal antiferromagnet YMnO$_3$, we have
found in the N\'{e}el phase a central peak at the 2D AFM zone
center which bears characteristics of the KT phase intrinsic to
the 2D XY spin systems. Understanding in detail how the N\'{e}el
and KT phases compete and change the nature of the
dynamic spin correlations would require further theoretical and 
experimental studies in the 2D XY spin systems.

\begin{acknowledgments}
 The authors thank K. Nho for providing us details of their theoretical calculations.
 Works at SPINS and DCS are partially supported by the NSF under DMR-9986442 and DMR-0086210, respectively.
 T.J.S. is supported by the Atomic Energy Division, MEXT, Japan.
\end{acknowledgments}



\begin{thebibliography}{19}
\bibitem[{pho()}]{phonon}
\bibinfo{note}{We rule out the phonon as the origin because the intensity does not follow the $Q^2$ behavior.}
\end{thebibliography}
\end{document}